\documentstyle[11pt,paspconf]{article}

\def\plotfiddle#1#2#3#4#5#6#7{\centering \leavevmode
\vbox to#2{\rule{0pt}{#2}}
\includegraphics{#1}}

\begin{document}

\title{{\it Ab Initio} Formation of Galaxies, Groups and Large-Scale 
Structure\footnotemark}\footnotetext{Invited 
talk at the IAU Colloquium 174 
``Small Galaxy Groups'', Turku, Finland, 13$^{\rm th}$-18$^{\rm th}$ 
June 1999.}

\author{Antonaldo Diaferio}
\affil{Max-Planck-Institut f\"ur Astrophysik, Karl-Schwarzschild-Str. 1, D-85740 
Garching, Germany}

\begin{abstract}
For the first time, the combination of 
semi-analytic modelling of galaxy formation and 
$N$-body simulations of cosmic structure formation 
enables us to model, at the same time, both 
the photometric and the clustering properties of galaxies.
Two Cold Dark Matter Universes provide 
a reasonable fit to the observed properties of galaxies, 
groups and clusters, including luminosities, colours,
density and velocity biases.
We show how the properties of galaxies and groups
on small scales are inextricably connected with the global
properties of the Universe.
\end{abstract}



\section{Introduction}

Groups of galaxies probe the intermediate scale between
galaxies and clusters. A satisfactory theory for the formation and the evolution
of these systems has always been difficult, because
the size of groups is small enough to require the modelling of the
properties and the internal structure of the
galaxy members and, at the same time, large enough to require the modelling
of the large scale structure sorrounding the groups.

For the last twenty years, $N$-body simulations 
have modelled groups with vacuum boundary conditions 
in order to resolve the internal kinematics of the galaxies
(e.g. Carnevali et al. 1981; Barnes 1985; Diaferio et 
al. 1993; Athanassoula et al. 1997). Large-scale
structure simulations have provided cosmological boundary 
conditions without resolving the galaxy internal properties
(e.g. Frederic 1995a,b; Nolthenius et al. 1997). 
Both sets of simulations have not modelled the 
physics of galaxy formation and evolution 
which provides observable quantities, such as luminosity and colour. 
However, modelling groups within a cosmological context with sufficient
resolution is extremely important, because
groups both trace the large-scale structure
of the Universe (e.g. Ramella et al. 1997, 1999) 
and are the sites of galaxy interactions 
(Hickson 1997; but see also e.g. Bettoni; Bosma; Rampazzo; Temporin, 
these proceedings). Galaxy interactions are connected with the formation of 
AGN's (Byrd, these proceedings), quasars (Cavaliere, these
proceedings) and multiple central black holes (Valtonen, these proceedings).
The importance of this large--small scale connection 
has been particularly important for successfully modelling 
the very existence of compact groups 
(Mamon 1986, 1989; Diaferio et al. 1994; Hernquist et al. 1995).
The enormous progress of the last few years in probing
the high-redshift Universe, which
shows a high degree of galaxy clustering (see e.g. 
Mazure \& Le F\`evre 1999), impels us to build
a cosmologically motivated theory of galaxy formation.

State-of-the-art $N$-body/hydrodynamic 
simulations have started only recently to barely reach the resolution required  
to follow the formation of galaxies within a cosmological context, even though
only some of the relevant physical processes are included (Pearce
et al. 1999).
For the time being, a more fruitful approach has
been the combination of semi-analytic modelling 
of galaxy formation with $N$-body simulations.
The semi-analytic approach
enables us to have full control of the relevant galaxy formation processes,
which occur on parsec or smaller scales, namely star formation, stellar evolution, 
and effects from supernova
explosions. The $N$-body simulations yield
the merging history of galaxies and their phase space position,
which are the result of group and cluster formation   
occurring on megaparsec
and larger scales.
By combining the two techniques, we are able to connect the stellar
population properties of galaxies with their clustering and kinematic properties.
This approach has already been particularly successful 
(Kauffmann et al. 1999a,b; Diaferio et al. 1999).

Here, I briefly review this technique (Section 2).
I then show some remarkable results which 
the self-consistency of the approach yields automatically:
the two component luminosity functions of groups (Section 3) and
the density and velocity biases of cluster galaxies
with differing colours
(Section 4). Extracting mock redshift surveys from
the simulation box also allows a direct comparison
with the observed large-scale distribution of galaxies
(Section 5).  
These results show the effectiveness of our
approach, and also the partial success of the
cosmological models.

\section{The Galaxy Formation Recipe}

According to standard inflationary models (e.g. Peacock 1999), 
large scale structure in the present day Universe originates
from primordial density perturbations amplified by gravitational
instability. These models also predict that most of the mass
in the Universe is non-baryonic and therefore ``dark''. 
Assuming a Cold Dark Matter (CDM) power spectrum of 
the perturbations (e.g. Efstathiou et al. 1992), the
clustering proceeds hierarchically: large perturbations collapse
first, forming small dark matter halos which aggregate to form larger
halos. Within these halos the baryonic gas 
cools, flows into the centre of the halo, and forms
stars which evolve and, eventually, some of them explode as supernovae.
The supernovae enrich the interstellar
gas with metals and reheat part of the gas.

We need two basic ingredients to implement this picture:
(1) the merging history of dark matter halos; and (2) recipes for the relevant
processes involving the baryonic matter.
Traditional semi-analytic models (e.g. Somerville \& Primack 1999 
and references therein) use Monte Carlo simulations
based on the extended Press \& Schechter (1974) formalism (e.g. Bower 1991)
to derive the merging tree of dark matter halos. Here, instead, we derive
the merger trees from
$N$-body simulations. Moreover, we identify the galaxy harbored by the halo with
the central dark matter particle of the halo:
the galaxy and the central particle share the same phase space
coordinates.
When two halos merge, the galaxy of the most massive halo ``jumps'' onto the
central particle of the resulting halo. The other galaxy still retains its
identity with its original dark matter particle, and becomes a ``satellite''
within the new halo. This approach makes the trajectory of
galaxies in phase space discontinous, but it assures a smooth evolution of the
luminosity and stellar mass of the galaxies.

We then implement the semi-analytic technique as follows.
We assume that the gas is in hydrostatic equilibrium within the dark matter halo.
We model both the baryonic and the dark matter components with 
truncated isothermal spheres.
The gas cools and flows on to the central particle of the halo
instantaneously, if the cooling time is shorter than the Hubble time at that epoch,
or at a given rate, otherwise.

The stellar mass $M_*$ increases accordingly to
\begin{equation}
{dM_*\over dt} = \alpha M_{\rm cold}\times 10{V_c\over R_{\rm vir}},
\end{equation}
where $M_{\rm cold}$ is the mass of cold gas, $R_{\rm vir}$ is the radius
of the halo of mass $M_{\rm vir}$, and $V_c^2=GM_{\rm vir}/R_{\rm vir}$ is
the halo circular velocity. The star formation efficiency $\alpha$ is
a free parameter.

Given the initial mass function for the stellar population (Scalo 1986), 
we can compute the number $\eta_{SN}$
of supernovae of Type II expected per unit mass. We neglect
supernovae of Type Ia. A fraction $\epsilon$, kept as a free parameter,
of the kinetic energy $E_{SN}$ ejected by each
supernova reheats part of the cold gas to the halo virial temperature. The mass
of the reheated gas over the time $\Delta t$ is
\begin{equation}
\Delta M_{\rm reheat} = \epsilon{dM_*\over dt}{4\eta_{SN} E_{SN}\over 3V_c^2}\Delta t.
\label{eq:reheat}
\end{equation}

Satellites merge with the central galaxy on a timescale set by dynamical
friction. We compute this timescale analytically. 
Our prescription does not allow for merging between satellites.
This shortcoming is partly responsible for the overproduction of
central bright galaxies, along with the 
fact that we do not implement a model of chemical
evolution (e.g. Kauffmann \& Charlot 1998),
which affects the gas cooling rate (see Section 3).
Luminosity evolution in different bands is computed with
stellar population synthesis models (Bruzual \& Charlot 1999).
Finally, galaxy luminosities are dimmed with an empirical prescription
for dust extinction (Cardelli et al. 1989). 

We apply our technique to two variants of a CDM Universe: 
a $\tau$CDM and a $\Lambda$CDM Universe.
A complete description of these models and our technique is
in Kauffmann et al. (1999a).

\section{The Luminosity Function of Galaxy Systems}

\begin{figure}
\plotfiddle{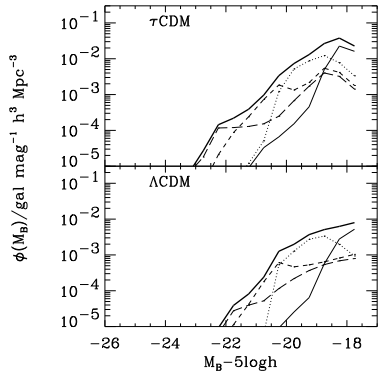}
           {0.20\vsize}              
           {90}                
           {30}                 
           {30}                 
           {130}               
           {-10}                
\caption{Blue-band luminosity function of galaxies within
systems of different total mass:
long-dashed line, $M>10^{14}h^{-1}M_\odot$; short-dashed line,
$10^{13}h^{-1}M_\odot<M\le10^{14}h^{-1}M_\odot$; dotted line,
$10^{12}h^{-1}M_\odot<M\le10^{13}h^{-1}M_\odot$; thin solid line,
$M\le10^{12}h^{-1}M_\odot$. The bold solid line is the total luminosity
function.}
\label{fig:LF}
\end{figure}

A scientific theory is required to provide testable
predictions. Sometimes, it also automatically provides 
explanations for phenomena for which the theory
was not originally constructed.

For example, consider 
the luminosity function of galaxies within different 
environments. It is well known that the usual
Schechter function (Schechter 1976) does not fit all
galaxy samples well. Many clusters (e.g. Biviano et al. 1995; 
Molinari et al. 1998; Trentham 1998), rich groups (Koranyi et al. 1998)
and even compact groups (Hunsberger et al. 1998; see, however,
e.g. de Carvalho et al. 1994 and Zepf et al. 1997) show a two component
luminosity function: a roughly Gaussian distribution
at the bright end, superimposed on a Schechter function
which dominates at the faint end. Because
ellipticals and S0's are generally more luminous than
spirals and irregulars, the Gaussian
component is mostly populated by early type galaxies, and
the Schechter component by late type galaxies.
Moreover, because this {\it bump} at the bright end peaks
at different magnitudes depending on the system considered,
when we consider a large galaxy sample, which includes different environments,
this feature disappears, and
a Schechter function provides a reasonably good fit
(e.g. Marzke et al. 1994).

Simulations of the formation of groups and clusters,
where galaxies are resolved and are allowed to merge,
predict a typical two component mass function at later times. 
This result has been well known since very early $N$-body
simulations (Aarseth \& Fall 1980; Roos \& Aarseth 1982: 
Cavaliere et al. 1991).
Thus, on the assumption that light is proportional to mass, galaxy
merging provides a natural explanation for the
bump observed in the luminosity function of galaxy
systems. The fact that this bump is not always present
clearly contains information on the formation history
of the galaxy system.

Our models, designed to reproduce the global properties
of galaxies and their large scale distribution rather
than the luminosity function of individual clusters,
also show this characteristic bump for massive systems
(long and short dashed lines in Fig. \ref{fig:LF}).

Apparently, the global luminosity function 
(bold solid line) shows no bump. Note that the global luminosity function 
is not a Schechter function. To obtain the Schechter form, it is sufficient to use
a dynamical friction prescription derived from high resolution $N$-body
simulations (Springel
et al. 1999), which automatically includes merging between satellites 
(Somerville \& Primack 1999), rather than our analytic prescription
(Section 2). The agreement with
the Schechter function can be improved further by modelling
the chemical enrichment of the interstellar
medium, which reduces the cooling efficiency at
early times.
 
\section{Morphology-Density Relation and Velocity Bias}

\begin{figure}
\plotfiddle{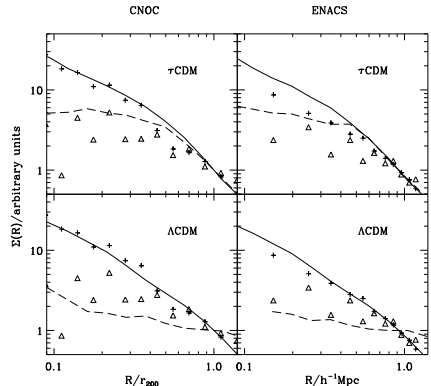}
           {0.20\vsize}              
           {90}                
           {30}                 
           {30}                 
           {130}               
           {-30}                
\caption{Surface number density profiles of galaxies within clusters of mass
$M>10^{14}h^{-1}M_\odot$ compared with observed cluster samples. 
Solid (dashed) lines show the profiles for red (blue) galaxies
in the models. Crosses and triangles are for red and blue galaxies in the
CNOC sample (left panels) and for non-emission-line galaxies
and emission-line galaxies for the ENACS sample (right panels).}
\label{fig:ENACS}
\end{figure}

The morphology-density relation has been known
since the very beginning of the systematic
investigation of galaxies (e.g. Hubble \& Humason 1931).
It has also been known that late type galaxies within clusters have
larger velocity dispersion than early type galaxies (e.g. Moss \& Dickens
1977). Modelling galaxy and large scale structure formation at
the same time provides a self-consistent way
of explaining these properties of cluster galaxies.

Fig. \ref{fig:ENACS} compares our models with 
the surface number density profiles of galaxy clusters. Profiles are for
blue and red galaxies (CNOC sample, Carlberg et al. 1997) and 
galaxies with emission and non-emission line spectra (ENACS,
Biviano et al. 1997). The agreement is satisfactory. In fact, the  
differences between models and observations are simply
due to the galaxy subsample definition: real
galaxies have been selected according to their spectral properties
which we do not model here.

The models can also account for the larger velocity
dispersion of blue galaxies compared to red galaxies. 
Blue galaxies are falling on to the
cluster for the first time and will become red later,
when they have exhausted their cold gas reservoirs.
Although this plausible explanation is well-known (e.g. Mohr et al. 1996),
our models yield this result self-consistently.
Moreover, our models show how this result depends on the cosmological
model (Diaferio et al. 1999, Diaferio 1999).

Our technique is therefore particularly promising.
For example, when simulations with better spatial
and force resolution become available, it will
be possible to investigate those kinematic properties 
of compact group members which, at present, seem 
difficult to interpret. For example, high velocity
galaxies (Sulentic, these proceedings) may simply be due to  
favourable, albeit rare, conditions of infall 
from the surroundings (Diaferio et al. 1994).

\section{Mock Redshift Surveys}

\begin{figure}
\plotfiddle{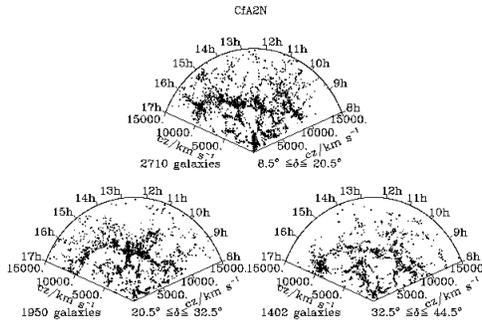}
           {0.2\vsize}              
           {90}                
           {30}                 
           {30}                 
           {130}               
           {-50}                
\caption{Galaxy distribution in the CfA2N catalogue 
projected on to three declination intervals.}
\label{fig:slice_cfa2n}
\end{figure}

\begin{figure}
\plotfiddle{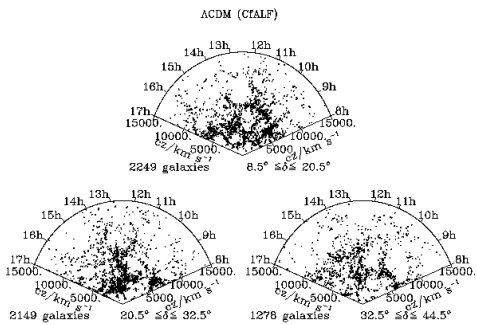}
           {0.29\vsize}              
           {90}                
           {30}                 
           {30}                 
           {130}               
           {-40}                
\caption{A mock catalogue extracted from the $\Lambda$CDM simulation box.}
\label{fig:slice_lcdm.cfa}
\end{figure}

Our simulation boxes have volumes that are comparable to
the volume probed by the northern region of the
Center for Astrophysics magnitude limited Redshift Survey (CfA2N hereafter; 
Geller \& Huchra 1989; 
Huchra et al. 1990; de Lapparent et al. 1991; Huchra et al. 1995;
Falco et al. 1999).
Therefore, we are able to compare our models
with a large portion of the Universe by 
analysing our simulation box the same way observers analyse their data.

We have compiled mock redshift surveys in order to have
a cluster as massive as Coma at the same
location as in the CfA2N (see Diaferio et al. 1999 for details).
Fig. \ref{fig:slice_cfa2n} shows the CfA2N and Fig. 
\ref{fig:slice_lcdm.cfa} shows a mock catalogue from one of our simulation boxes. 
The large scale structure in our simulation is not
as sharply defined as in the real Universe: for example 
no Great Wall (Geller \& Huchra 1989) is present in the model. 
Schmalzing \& Diaferio (1999) have quantified these 
topological differences on large
scale using Minkowski functionals.

Further constraints on the model of galaxy and
structure formation come from clustering
on smaller scales, in particular from the
properties of galaxy groups (Ramella et al. 1997, 1999; Tucker,
these proceedings).

Catalogues of groups are usually compiled using a 
friends-of-friends algorithm on redshift data (Huchra \& Geller 1982).
However, in these redshift surveys, we do not know the true distance to a galaxy; 
we only know its redshift, which
also includes the galaxy peculiar velocity.
Therefore, the average properties of groups
extracted from a redshift survey do not necessarily agree with 
those of groups extracted from an ideal survey in configuration
space. 

$N$-body simulations allow us to quantify this difference.
Although previous analyses suggest that the differences are small 
(e.g. Nolthenius \& White 1987;
Moore et al. 1993; Frederic 1995a, 1995b; Nolthenius et al. 1997),
these analyses do not include the physics of galaxy formation. 

\begin{figure}
\plotfiddle{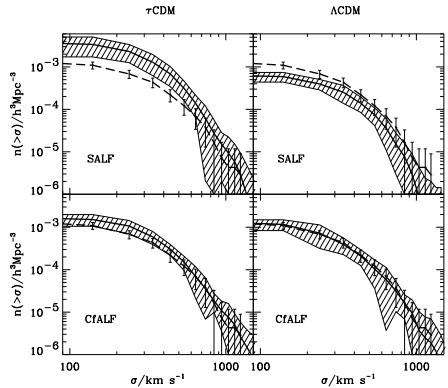}
           {0.20\vsize}              
           {90}                
           {30}                 
           {30}                 
           {130}               
           {-30}                
\caption{Group abundance by velocity dispersion $\sigma$. Bold lines are the mean
number densities averaged over an ensamble of ten mock catalogues. Shaded 
areas show
the 3-$\sigma$ deviations. Dashed lines are for the CfA2N groups.
Error bars on the CfA2N curves are Poisson 3-$\sigma$ deviations.
Top and bottom panels are for two different model luminosity functions
(see Diaferio et al. 1999).}
\label{fig:gr_nd_sigma}
\end{figure}

On the other hand, we are able to compile redshift surveys that 
take into account the luminosities of galaxies, which result from their
merging and stellar evolution histories.
We find that typical parameters of the 
friends-of-friends algorithm yield
kinematic properties which are in agreement with
both those of groups selected from configuration
space and those of the CfA2N groups (Fig. \ref{fig:gr_nd_sigma}).
However, 40\% of the triplets selected in
redshift space are, in fact, unrelated galaxies
which are {\it not} bound systems in configuration
space. 

Some of the differences between groups in 
the models and in the CfA2N is due to the differing
distribution of galaxies on large scale.
An $N$-body simulation constrained to have
the same large scale structure as the local
Universe (Mathis et al. 1999) will be valuable for further constraining 
the galaxy formation recipes using the properties of groups.

\section{Conclusion}

Combining semi-analytic modelling with $N$-body
simulations is an extremely powerful tool for
investigating galaxy and group formation within
a self-consistent cosmological framework.
$N$-body simulations with better spatial and force
resolution than the simulations presented here
will be ideal for understanding
the properties of compact groups, their different morphology and
environment (e.g. de Carvalho, these proceedings), their
abundance in the local Universe and at high redshift.

Groups selected objectively from two-dimensional
information (Prandoni et al. 1994), with follow-up
redshift measurements (Iovino, these proceedings),
or directly selected in redshift space (Barton et al. 1996; Allam,
these proceedings), can provide a wealth of
information on both cosmology and galaxy formation processes.

The physics of the intergalactic medium
should also be included in our approach
(Cavaliere et al. 1998; Cavaliere, these
proceedings): both HI (e.g. Combes; Verdes-Montenegro; Sancisi,
these proceedings) and X-ray observations (e.g. Mulchaey;
Ponman, these proceedings)
can show clear signatures of galaxy interactions
and tell us about the dynamical state of the group (Diaferio et al. 1995).

To simulate the evolution of a group of galaxies, we can 
extract the initial conditions and the external 
tidal field from a large-scale $N$-body simulation.
This strategy has already been extremely successful
in simulating galaxy clusters (Tormen et al. 1997; Springel et al. 1999).
Thus, high-resolution $N$-body/hydrodynamic simulations will enable us 
to predict the properties of the galaxies and the intergalactic medium of 
individual groups within a full cosmological context.

\acknowledgments

The $N$-body simulations were carried
out at the Computer Center of the Max-Planck Society
in Garching and at the EPPC in Edinburgh, as part of the Virgo Consortium project.
I thank J\"org Colberg, Guinevere Kauffmann and Simon White 
for permission to reproduce results from our joint research programme 
within the German Israeli Foundation (GIF) collaboration.
I thank Margaret Geller and John Huchra for allowing 
the comparison of the CfA2N data directly with the simulations,
and Ravi Sheth for a careful reading of the manuscript.
I thank Mauri Valtonen for the invitation to this meeting, and 
him and the other organizers for
a stimulating conference and a delightful time in Turku.
Special thanks to Reinaldo de Carvalho and Giorgio Palumbo for lovely discussions
and unexpected lightening endings of some Finnish nights.

\end{document}